# A model that solves to the wave-particle duality paradox

Eduardo V. Flores[1]

**Abstract**
Bohmian mechanics solves the wave-particle duality paradox by introducing the concept of a physical particle that is always point-like and a separate wavefunction with some sort of physical reality. However, this model has not been satisfactorily extended to relativistic levels. Here we introduce a model of permanent point-like particles that works at any energy level. Our model seems to have the benefits of Bohmian mechanics without its shortcomings. We propose an experiment for which the standard interpretation of quantum mechanics and our model make different predictions.

**1. Introduction**

Quantum physics numerical predictions are outstanding at all the available energy levels. However, the same is not true about the interpretation of quantum physics [1-5]. Standard quantum theory contains a paradox known as the wave-particle duality paradox, that is, how the same object could sometimes be extended so as to produce wave interference and instantaneously it could become point-like such as a detected particle on a screen. This paradox might be symptomatic of a theory with intrinsic problems or a theory with an incorrect interpretation of its results. The mathematical success of quantum mechanics points to a problem with the interpretation of the theory.

A model known as Bohmian mechanics originally proposed by Louis De Broglie and then expanded by David Bohm has no wave-particle duality paradox while the numerical results are similar to the standard quantum theory at least at the non-relativistic level. In Bohmian mechanics particle and wave are different objects, therefore, there is no duality paradox. In this theory the particle is always a point-like physical object. The wave is a separate object that guides the motion of the particle. John Bell defended this model [6]:

*While the founding fathers agonized over the question 'particle' or 'wave', de Broglie in 1925 proposed the obvious answer 'particle' and 'wave'. Is it not clear from the smallness of the scintillation on the screen that we have to do with a particle? And is it not clear, from the diffraction and interference patterns, that the motion of the particle is directed by a wave? De Broglie showed in detail how the motion of a particle, passing through just one of two holes in screen, could be influenced by waves propagating through both holes. And so influenced that the particle does not go where the waves cancel out, but is attracted to where they cooperate. This idea seems to me so natural and simple, to resolve the wave-particle dilemma in such a clear and ordinary way, that it is a great mystery to me that it was so generally ignored.*

"Bohmian mechanics is a theory about point particles moving along trajectories. It has the property that in a world governed by Bohmian mechanics observers see the same statistics for experimental results as predicted by quantum mechanics. Bohmian mechanics thus provides an explanation of quantum mechanics" [7]. Bohmian mechanics is a quantum theory describing the motion of point-like particles with definite trajectory [8]. We see that researchers working on Bohmian mechanics claim that this theory provides a consistent resolution of all paradoxes of quantum mechanics, in particular of the so-called measurement problem [7-11]. In Bohmian mechanics the position of a particle changes according to an equation of motion known as the guiding equation. For instance, for a non-relativistic particle of mass $m$ without spin the guiding equation is

$$\frac{dQ}{dt} = \frac{\hbar}{m} \operatorname{Im} \frac{\nabla \psi}{\psi} \tag{1}$$

where $Q$ is the particle coordinate, $\psi$ is the wavefunction, and $t$ is time. Bohmian mechanics is a fully deterministic theory of particles in motion, but a motion of a profoundly non-classical, non-Newtonian sort [9]. As an example we consider the case of a particle in a box. The wavefunction for the particle in box has no imaginary part, thus, according to Eq. 1 the particle is at rest no matter what energy it has; a similar thing happens to the electron in the ground state of hydrogen [10]. The generalization of Bohmian mechanics to the relativistic level has not been successful. Bohmian mechanics has serious problems with Lorentz invariance. The problems of Bohmian mechanics stem from the use of the guiding equation, Eq. 1 [11].

---

[1] Department of Physics & Astronomy, Rowan University, 201 Mullica Hill Rd., Glassboro, NJ 08028. E-mail: flores@rowan.edu

If a theory of point-like particles at all times is going to succeed, at all levels of energy, a new approach is needed to obtain particle trajectory. We must avoid using of Eq. 1 to obtain particle trajectory. We notice that a point-like particle is detected by its energy, momentum, charge, etc. These conserved quantities are not arbitrary but follow precise rules. In addition, when we follow the energy, momentum and charge we are really following the particle. Therefore, the best way to obtain particle trajectory is by applying the conservation laws. Our technique to obtain particle trajectory is by direct measurement of partial particle trajectory plus extrapolation of missing sections by applying the conservation laws. This technique to determine particle trajectory is heavily used in experimental particle physics [12]. Thus, we can determine particle trajectory for different particles from detectors all the way back to the original particle collision. Using this technique, particles such as the Higgs boson and weak vector bosons have been discovered. At the theoretical level, John Wheeler uses this technique to determine particle trajectory in his delayed choice experiment [2]. A great advantage of using the conservation laws to obtain particle trajectory is that it applies equally well at relativistic and non-relativistic levels.

We may investigate properties of a quantum theory of point-like particles by considering a photon that enters an ideal 50:50 beam splitter. We assume that a wave packet with a single point-like particle enters a beam splitter as in Fig. 1. The incident wave packet obeys Maxwell's equations and boundary conditions; thus, it splits into two smaller wave packets. However, the particle, which enters with the incident wave packet, must take one of the two available paths. We note that empty and occupied wave packets would propagate in vacuum until they dissipate.

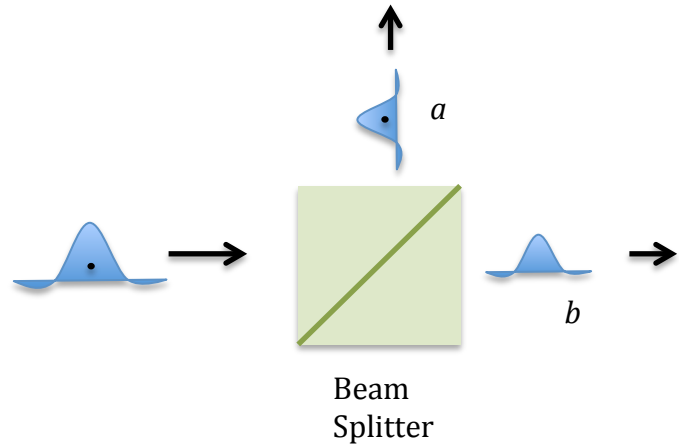

**Fig. 1| A particle and a field through a beam splitter.** A wave-packet enters the beam splitter and breaks into two smaller wave-packets. The incident particle, shown as a dot in picture, emerges along path *a* accompanied by a wave packet. The wave packet in path *b* is empty.

By definition, the two outgoing wave packets exiting the 50:50 beam splitters must be identical. Thus, the wave packets are only affected by the boundary condition, which in this case is the beam splitter. The wave packets propagate according Maxwell's equations. The actual path the particle takes does not affect the wave packets. We also notice that the location of the particle within the wave packet makes no difference to the fields.

Here is another major difference between our model and Bohmian mechanics. In our model the field does not carry significant or measurable amount of energy, momentum or angular momentum. All of the conserved quantities are carried by the particle. Therefore, the field cannot drive the particle. In Bohmian mechanics the field drives the particle, as it is evident by looking at the guiding equation, Eq. 1. The guiding equation shows that the particle trajectory is directly influenced by the state of the field. In our model we consider the fields as only providing possibilities for particle motion and interaction. The fields present possibilities to the particle but the conservation laws determine where the particle actually goes. In classical physics the particle has no options; its motion is deterministic. The quantum field shows a much richer landscape for the particle; this is a dynamical landscape as the fields are not static but always changing affected by changes in boundary conditions and interaction with other fields.

In high-energy physics we find a useful example of the role of the field in our model. Quantum fields have on mass-shell and off mass-shell states. On mass-shell states represent the presence of physically real particles. However, according to Veltman [13]: "Particles off mass-shell, virtual particles, are parts of diagrams, and we may even have some intuitive feeling about them, but we should never make the mistake of treating them as real particles." We note that even though virtual particles contribute to physical effects, they, themselves, are not real. As an example, we consider a high-energy photon propagating in vacuum; the photon field contains a virtual electron-positron pair. However, if the photon were to spontaneously decay into an electron and a positron there would be a violation of momentum conservation. Therefore, the electron and positron cannot be spontaneously formed. What is needed to make this virtual pair real is the presence of an external object, such as the nucleus of an atom to provide the required momentum. Thus, the field presents possibilities to the particle; however, only when the conservation laws are fulfilled the possibilities could become real.

## 2. The model

Our model is based on the following assumptions:

I. *The particle is always a point-like obj*ect
II. *The particle carries conserved quantities*
III. *Particle trajectory is determined by the conservation laws*
IV. *For every type of elementary particle there is a quantum field*
V. *The field reveals the possibilities for particle motion and interaction*
VI. *The field develops according to field equations and boundary conditions*
VII. *The location of the particle does not affect the development of the field*

## 3. A prediction to be tested

Some years ago Afshar *et al* claimed to have simultaneously observed physically real (in the classical sense) particle-like and wave-like properties of photons beyond the limit imposed by complementarity [14]. The setup in Fig. 2 is an optical equivalent of the experimental setup of Afshar *et al.* Two beams meet and interfere constructively and destructively with high visibility, $V \approx 1$, as shown by the minimal changes in the readings of end detectors when a wire grid is placed at the center of the dark fringes. Several authors [15-17] have shown that if the wires are thin enough and placed at the center of dark fringes the beams are hardly deflected and most photons that hit detector 1 (2) come from beam 1 (2); this results in a high value for the which-way information parameter $K$. We note that $K$ the measures the degree to which particle trajectory is determined.

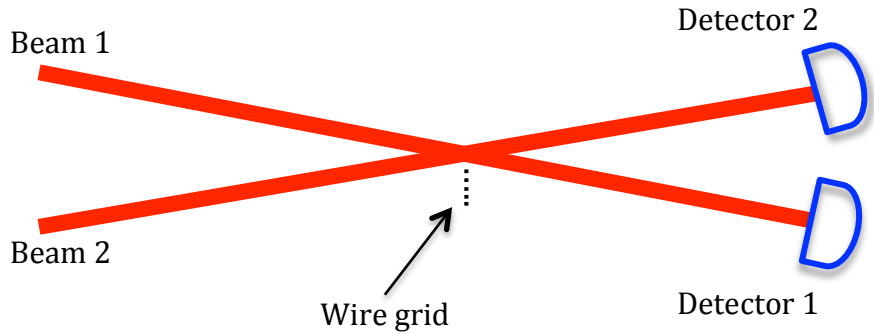

**Fig. 2| Experiment with two laser beams.** Two laser beams cross and impinge upon different detectors that identify the origin of each photon, providing which-way information. A wire grid is scanned across the beam intersection as a probe of an interference pattern. If the readings at end detectors significantly decrease then the wires at the center of bright fringes; when the reading are unchanged as compared to the case without the wire grid then the wires at the center of dark fringes.

Afshar *et al* results violate the complementarity inequality also known as the Englert-Greenberger inequality [18,19]

$$K^2 + V^2 \leq 1 \quad (2)$$

Researchers responding to these paradoxical findings have pointed out potential problems with the identification of which-way information [17,20]; others have obtained a visibility low enough to preserve the inequality in Eq. 2 [15,16]. However, these resolutions have been challenged [17,21]. Except for the particular case when the beams are entangled the paradoxical findings of Afshar *et al* stand [17].

At the experimental level, Jacques *et al* [16] put forward an ambitious attempt to resolve the paradox posed by the experimental results of Afshar *et al*. Their setup is similar to the setup in Fig. 2. In their experiment they use a train of single photon pulses that is split into two beams that interfere and separate. Their theoretical approach is based on Fraunhofer diffraction. They obtain an expression for the which-way information consistent with other workers in the field [15,17]. Their experimental results agree with their calculations for the signal at the detectors. However, it has been pointed out that Jacques *et al* value for the which-way information is not valid within the standard interpretation of quantum mechanics [17]. The problem with their experiment is their use of a true single photon that is split, by a partially reflective beamsplitter, into the two beams required for the experiment. When a single photon goes through the beamsplitter the state of the photon is described by $|1\rangle_a|0\rangle_b + |0\rangle_a|1\rangle_b$, where $a$ and $b$ represent the two beams; thus, the photon is entangled [22]. According to the standard interpretation of quantum mechanics an entangled photon is equally present in both beams, thus, it has no which-way information ($K = 0$) [17,22]. This fact does not allow us to use the experimental results of Jacques *et al* to refute the results of Afshar *et al*.

We notice that if an opaque screen were placed where the beams cross in the setup in Fig. 2 we would see an interference pattern on the screen, even for the case when the two beams are generated by independent lasers. In fact, this is the experimental setup for the Pfleegor and Mandel experiment. In 1967 Pfleegor and Mandel reported [3] that they observed interference of single photons from independent photon sources. Pfleegor and Mandel associate the formation of the interference pattern with the intrinsic uncertainty from which of the two sources the photon comes from [3]. Or as R. Loudon put it [22]: "The interference occurs between the probability amplitudes that the detected photon was emitted from one source or the other." This means that if we were to somehow discover the origin of each photon then the interference fringes would disappear. This interpretation is based on the standard interpretation of quantum mechanics. Thus, we can make a prediction for the setup in Fig. 2 when using independent photon beams.

Here is what the standard interpretation of quantum mechanics would predict for the setup in Fig. 2 with independent photon beams [3,22]. We could pick the wires to be thin enough so that the beams are hardly disturbed by them. This means that when a photon hits detector 1 (2) it came from beam 1 (2) with a high level of certainty. Thus, the degree of which-way information could be as high as we wish, $K \approx 1$. In this case, since we know the origin of each photon we should not get interference fringes. In fact, the inequality in Eq. 2 implies that the visibility should be nearly zero, $V \approx 0$, when the which-way information is nearly one, $K \approx 1$. Thus, if the wires are scanned across the beam intersection we should see no significant increases or decreases in the signal at end detectors showing that there are no sharp interference fringes at the intersection.

On the other hand, according to our model, the fields from each beam that meet at the intersection should interfere constructively and destructively. However, this is just for the fields, thus, it is at the level of possibilities only. When there is no wire grid a particle in a beam would cross the beam intersection undeflected. This is so because for a photon to be deflected it takes momentum. If momentum were lacking even if the fields show interference fringes the particle would not be deflected. These particles have full which-way information but do not form an interference pattern due to the lack of momentum conservation. Thus, the inequality in Eq. 2 is preserved.

If thin wires were located at the center of dark fringes we predict that the readings at the end detectors would be hardly different from the case without the wire grid. This means that now the photons that come close to the wires are deflected, as the fields indicate, to avoid the wires. This is so because the wires can provide the momentum necessary to deflect the photons. These photons have been deflected and carry high visibility, $V \approx 1$. However, the uncertainty principle shows that these photons, deflected by very thin wires, would get a momentum uncertainty large enough to wash out information about their origin. Photons that pass far from the wires would not be deflected thus they have no visibility, $V \approx 0$, but keep high which-way information, $K \approx 1$. Thus, the inequality in Eq. 2 would be preserved in all cases. We conclude that our model predicts the observation of sharp interference fringes for the setup in Fig. 2 with independent beams.

**4. Conclusions**

We propose a model where the particle is a permanent point like object that carries energy, momentum, charge, etc. The wave is a solution to a field equation in quantum physics. Our model is an alternative to Bohmian mechanics. What makes our model more attractive than Bohmian mechanics is that to obtain particle trajectory we just apply the conservation laws. In Bohmian mechanics, to obtain particle trajectory a new equation has to be added to the theory; this equation encounters difficulties with Lorentz invariance and produces some non-intuitive results not yet confirmed by experiment. We find in our model a successful theory of quantum physics not only for numerical purposes but also for interpretational purposes. In the future, we would like to investigate more about the nature of the fields. We would like to address how other quantum paradoxes may be dealt within our model.

We predict that for the setup in Fig. 2, with two independent photon beams that cross and separate, there would be interference fringes with high visibility. This would happen even though the which-way information is high; however, our model shows that there would be no violation of the complementarity inequality, Eq. 2. On the other hand, following the work of Pfleegor and Mandel and the comments of R. Loudon we conclude that the prediction of the standard interpretation of quantum mechanics for the same setup would show that there are no sharp interference fringes. This is due to the high level of certainty the setup in Fig. 2 provides about the origin of the photons. We look forward to seeing the results of an experiment with independent photon beams using the setup in Fig. 2.